\documentclass{emulateapj}
\usepackage{apjfonts}

\usepackage{mathrsfs}
\usepackage{CJK}

\usepackage{amsmath}

\usepackage{graphics}

\newcommand{\bd}{\begin{displaymath}}
\newcommand{\ed}{\end{displaymath}}

\shorttitle{The outflows accelerated by the magnetic fields and
radiation force}

\shortauthors{Xinwu Cao}

\begin{document}
\title{The outflows accelerated by the magnetic fields and radiation force of accretion
disks}
\author{Xinwu Cao}
\affil{Key Laboratory for Research in Galaxies and Cosmology,
Shanghai Astronomical Observatory, Chinese Academy of Sciences,\\ 80
Nandan Road, Shanghai, 200030, China; cxw@shao.ac.cn}

\slugcomment{accepted by ApJ}

\begin{abstract}
The inner region of a luminous accretion disk is radiation pressure
dominated. We estimate the surface temperature of a radiation
pressure dominated accretion disk, $\Theta=c_{\rm
s}^2/r^2\Omega_{\rm K}^2\ll (H/r)^2$, which is significantly lower
than that of a gas pressure dominated disk, $\Theta\sim (H/r)^2$.
This means that the outflow can be launched magnetically from the
photosphere of the radiation pressure dominate disk only if the
effective potential barrier along the magnetic field line is
extremely shallow or no potential barrier is present. For the latter
case, the slow sonic point in the outflow may probably be in the
disk, which leads to a slow circular dense flow above the disk. This
implies that hot gas (probably in the corona) is necessary for
launching an outflow from the radiation pressure dominated disk,
which provides a natural explanation on the observational evidence
that the relativistic jets are related to hot plasma in some X-ray
binaries and active galactic nuclei. We investigate the outflows
accelerated from the hot corona above the disk by the magnetic field
and radiation force of the accretion disk. We find that, with the
help of the radiation force, the mass loss rate in the outflow is
high, which leads to a slow outflow. This may be the reason why the
jets in radio-loud narrow-line Seyfert galaxies are in general mild
relativistic compared with those in blazars.
\end{abstract}

\keywords{accretion, accretion disks, galaxies: jets, magnetic
fields, galaxies: active}

\section{Introduction}

The acceleration of the gas by the magnetic field lines threading
the rotating disks is considered as a promising explanation for
jets/outflows observed in different types of the sources, i.e.,
young stellar objects, X-ray binaries, and active galactic nuclei
(AGNs) \citep*[see reviews
in][]{1996epbs.conf..249S,2000prpl.conf..759K,2007prpl.conf..277P,2010LNP...794..233S}.
In this model, a fraction of gas is driven form the disk along the
field line co-rotating with the disk by the centrifugal force
\citep{1982MNRAS.199..883B}. The jets/outflows are dominantly
powered by the rotational kinetic energy of the disk through the
ordered large scale field threading the accretion disk.

It has been also suggested that the outflow can be accelerated by
the radiation pressure of the disk
\citep*[e.g.,][]{1977A&A....59..111B,1985ApJ...294...96S,1995ApJ...451..498M}.
This scenario is attractive especially for the jets in the sources
accreting at high mass rates. Indeed, jets have been observed in the
black hole systems accreting at high rates, such as, some
narrow-line Seyfert I galaxies (NLS1s), young radio galaxies, and
microquasars
\citep*[e.g.,][]{2003ApJ...584..147Z,2006PASJ...58..777D,2008ApJ...685..801Y,2010AJ....139.2612G,2009ApJ...698..840C,2009ApJ...701L..95W,2004MNRAS.355.1105F}.
The radiation pressure may also play an important role in the
magnetically driven outflows especially from luminous accretion
disks. \citet{2000ApJ...538..684P} performed numerical simulations
on the radiation-driven outflows from a luminous Keplerian accretion
disk threaded by a strong large-scale ordered magnetic field. It has
been found that the radiation force is essential in driving outflows
from the disk if the thermal energy of the gas is low or the field
lines make an angle greater than 60$^\circ$ with respect to the disk
midplane. Strictly speaking, the circular motion of the gas in the
accretion disk always deviates from Keplerian motion in the presence
of a large scale magnetic field, which usually exerts a radial force
against the gravitation of the central object. Therefore, the
accretion disk threaded by ordered magnetic field lines is therefore
always sub-Keplerian
\citep*{1998ApJ...499..329O,2001ApJ...553..158O,2002A&A...385..289C,2012MNRAS.426.2813C}.
The radiation-magnetohydrodynamic simulations have been carried out
on the global structure of the accretion disks and outflows around
black holes
\citep*[e.g.,][]{2007A&A...474....1M,2009ApJ...706..497M,2009MNRAS.397.2087M,2010PASJ...62L..43T,2011ApJ...736....2O,2011ApJ...742...56V}.
It is found that the magnetic force, together with the radiation
force exerted by accretion disks, can efficiently drive outflows
from luminous accretion disks. \citet{2012MNRAS.426.2813C}
investigated the launching condition for the cold outflows driven by
the magnetic field and radiation force of an accretion disk, in
which the disk is sub-Keplerian due to the force exerted by the
magnetic field. It has been found that the force exerted by the
radiation from the disk does help to launch the outflow. The cold
gas can be driven from the disk surface even if the field line is
inclined at angle greater than $60^\circ$ with respect to the disk
surface, which is obviously different from the pure magnetically
driven cold outflow \citep*{1982MNRAS.199..883B}.

In this work, we explore the outflow accelerated from the hot corona
above the disk by the magnetic field and radiation force of the
disk. In our model, the disk structure, especially the circular
motion velocity of the gases in the disk, is altered in the presence
of a strong large-scale ordered magnetic field threading the disk.
We describe the model and results in Sections 2 and 3, and the
discussion is in Sections 4. The final section contains a summary.

\vskip 1cm
\section{Model}

We consider the outflow accelerated by the large scale magnetic
field and the radiation force of a radiation pressure pressure
dominated accretion disk. The rotational velocity of the gas in the
disk is sub-Keplerian in the presence of a magnetic field, due to a
radial magnetic force against the gravity of the central object. The
outflow is fed by the gas at the disk surface. The properties of the
transition region between the disk and outflow are still quite
uncertain, though some efforts have been devoted on this issue
\citep*[e.g.,][]{1998ApJ...499..329O,2001ApJ...553..158O}. In this
work, we consider the outflows fed by the hot corona above the disk,
and the temperature and density of the corona are described as the
input model parameters. A large scale magnetic field threading the
disk is a key ingredient in our model calculations, however, the
origin of such a field is still not well understood. The field can
be advected inwards by the accretion matter from the interstellar
medium or a companion star
\citep*[][]{1974Ap&SS..28...45B,1976Ap&SS..42..401B}. The inward
advection of the field lines is balanced by the outward movement of
field lines caused by magnetic diffusion
\citep*[e.g.,][]{1989ASSL..156...99V,1994MNRAS.267..235L}. In
principle, the field configuration can be calculated if the
structure of the disk is known. \citet{1994MNRAS.267..235L} found
that the advection of the field in a conventional viscously driven
thin accretion disk is rather inefficient. It was argued that the
accretion velocity of the gas in the region away from the midplane
of the disk can be larger than that at the midplane of the disk,
which makes the field dragged in by the accretion disk more
efficiently
\citep*[][]{2009ApJ...701..885L,2012MNRAS.424.2097G,2013MNRAS.430..822G}.
\citet{2013ApJ...765..149C}'s calculations show that the external
field can be advected inwards efficiently if the angular momentum of
the disk is predominately removed by the outflows. {An alternative
for the origin of magnetic fields is the dynamo processes in the
disks, and the outflows can be driven by the dynamo generated
magnetic fields of the disks \citep*[][]{
2003A&A...398..825V,2003MNRAS.345..123C}.} To avoid the complexity
of the detailed physics of the disk field formation, we use an
analytic magnetic field configuration of an accretion disk given in
\citet{1994A&A...287...80C} to investigate the dynamics of the
outflow driven by the magnetic field and radiation pressure of the
disk.

\subsection{Structure of a radiation pressure dominated accretion
disk}

The structure of a radiation pressure dominated accretion disk with
a large scale magnetic field has been studied in detail by
\citet{2012MNRAS.426.2813C}. We briefly summarize the main results
in this subsection.

The half-thickness $H_{\rm d}$ of a radiation pressure dominated
accretion disk is estimated by assuming the vertical component of
gravity to be balanced with the radiation force and the vertical
component magnetic force at the disk surface. {The curvature of the
field line at the disk surface is usually very small, and the
vertical magnetic force can be neglected compared with the radiation
force at $z=H_{\rm d}$. The disk thickness can be calculated with}
\begin{equation}
{\frac {GMH_{\rm d}}{(r_{\rm i}^2+H_{\rm d}^2)^{3/2}}}={\frac
{f_{\rm rad}\kappa_{\rm T}}{c}}, \label{vertical_1}
\end{equation}
where $f_{\rm rad}$ is the flux from the unit surface area of the
disk, and $\kappa_{\rm T}$ is the Thompson scattering cross-section.

The angular velocity $\Omega$ of the disk is calculated with
\begin{equation}
r_{\rm i}\Omega_{\rm K}^2-r_{\rm i}\Omega^2={\frac
{B_r^{S}B_{z}}{2\pi\Sigma_{\rm d}r_{\rm i}}}={\frac
{B_{z}^2}{2\pi\Sigma_{\rm d}r_{\rm i}\kappa_0}}, \label{Omega_1}
\end{equation}
where $B_r^{\rm S}$ and $B_z$ are the radial and vertical components
of the field at the disk surface, respectively,
$\kappa_0=B_z/B_r^{\rm S}$, and $\Sigma_{\rm d}$ is the surface
density of the disk. The footpoint of the field line at the disk
surface is located at $r=r_{\rm i}$ and $z=H_{\rm d}$.

The pressure at the mid-plane of an accretion disk is
\begin{equation}
p_{\rm d}={\frac {4\sigma}{3c}}T_{\rm c}^4,\label{p_d_1}
\end{equation}
where $T_{\rm c}$ is the central temperature of the disk. The flux
radiated from the unit area of the disk surface is related to the
temperature and density of the disk by
\begin{equation}
f_{\rm rad}={\frac {8\sigma T_{\rm c}^4}{3\Sigma_{\rm d}\kappa_{\rm
T}}}.\label{f_rad_1}
\end{equation}

Substituting Equations (\ref{vertical_1}), (\ref{p_d_1}), and
(\ref{f_rad_1}) into Equation (\ref{Omega_1}), we find
\begin{equation}
1-\tilde{\Omega}^2={\frac {2\beta\tilde{H}_{\rm
d}}{\kappa_0(1+\tilde{H}_{\rm d}^2)^{3/2}}}, \label{Omega_2}
\end{equation}
where the dimensionless quantities $\tilde{\Omega}$, $\tilde{H}_{\rm
d}$, and $\beta$, are defined as
\begin{equation}
\tilde{\Omega}={\frac {\Omega}{\Omega_{\rm K}}};~~~~~\Omega_{\rm
K}=\left({\frac {GM}{r_{\rm i}^3}}\right)^{1/2};~~~~~\tilde{H}_{\rm
d}={\frac {H}{r_{\rm i}}};~~~~~\beta={\frac {B_z^2}{8\pi}}/p_{\rm
d}.\label{quantities}
\end{equation}

\subsection{Magnetic field configuration}

We use a magnetic field configuration with
\begin{equation}
B_z(r,~z=0)=\left[1+\left({\frac
{r}{r_0}}\right)^2\right]^{-1/2},\label{b_z_0}
\end{equation}
at the mid-plane of the disk \citep*[][]{1994A&A...287...80C}, which
is similar to the self-similar disk given in
\citet{1982MNRAS.199..883B} \citep*[see][for the detailed
discussion]{1996epbs.conf..249S}. The stream function of the
potential field in the space above/below the disk satisfying the
boundary condition (\ref{b_z_0}) is given by
\begin{equation}
\Phi=\left[\left({\frac {r}{r_0}}\right)^2+\left(1+\left|{\frac
{z}{r_0}}\right|\right)^2 \right]^{1/2}-\left(1+\left|{\frac
{z}{r_0}}\right|\right). \label{phi_1}
\end{equation}
It satisfies the conditions $\bigtriangledown\times \bf{B}=0$ and
$\bigtriangledown\cdot\bf{B}=0$, and it is used in this work to
study the dynamics of the outflow above the disk. The components of
the field in the space above the disk are given by
\begin{equation}
B_z(r,~z)={\frac {1}{r}}{\frac {\partial \Phi}{\partial r}},
\label{b_z_1}
\end{equation}
and
\begin{equation}
B_r(r,~z)=-{\frac {1}{r}}{\frac {\partial \Phi}{\partial z}}.
\label{b_r_1}
\end{equation}

{In principle, the radial magnetic field will be sheared into
azimuthal field by the differential rotation of the gas in the disk,
which leads to magnetorotational instability (MRI) and the
MRI-driven turbulence
\citep*[][]{1991ApJ...376..214B,1998RvMP...70....1B}. This is the
most promising mechanism for angular momentum transportation in
accretion disks. In this work, we consider the properties of the
disk as the boundary conditions for driving the outflows, and
therefore we do not need to consider the detailed physics of the
azimuthal field in the disk caused by the differential rotation.}

\subsection{Radiation of the accretion disk}

The radiation force exerted by the disk is in $z$-direction at the
disk surface, which can be determined locally, $\mathscr{F}_{\rm
rad}(z=H_{\rm d})=f_{\rm rad}\kappa_{\rm T}\rho_{\rm i}/c$ , where
$\rho_{\rm i}$ is the density of the gas at the disk surface. In the
space above the disk, the radiation force exserted by the disk has
to be calculated by integrating over the contribution from the whole
disk \citep*{1977A&A....59..111B}.

The $r$ and $z$-components of the radiation force exerted on the gas
element with density $\rho$ at $(r,~z)$ above the disk are
\begin{displaymath}
\mathscr{F}_{{\rm rad},r}(r,~z)={\frac {\kappa_{\rm
T}\rho}{c}}~~~~~~~~~~~~~~~~~~~~~~~~~~~~~~~~~~~~~~~~~~~~~~~~~~~~~~~~~~~~~~~~~
\end{displaymath}
\begin{equation}
\times\int\limits_{r^\prime}\int\limits_{\phi^\prime}{\frac {f_{\rm
rad}(r^\prime)\mu^\prime(1-{\mu^\prime}^2)^{1/2}(r-r^\prime\cos\phi^\prime)}{\pi
l(h^2+l^2)}}r^\prime dr^\prime d\phi^\prime, \label{force_rad_r1}
\end{equation}
and
\begin{equation}
\mathscr{F}_{{\rm rad},z}(r,~z)={\frac {\kappa_{\rm
T}\rho}{c}}\int\limits_{r^\prime}\int\limits_{\phi^\prime} {\frac
{f_{\rm rad}(r^\prime)h^2}{\pi (h^2+l^2)^2}}r^\prime dr^\prime
d\phi^\prime, \label{force_rad_z1}
\end{equation}
where
\begin{displaymath}
h=z-\tilde{H}_{\rm d}r,
\end{displaymath}
\begin{displaymath}
l=[(r-r^\prime\cos\phi^\prime)^2+{r^\prime}^2\sin^2\phi^\prime]^{1/2},
\end{displaymath}
and
\begin{displaymath}
\mu^\prime={\frac {h}{(h^2+l^2)^{1/2}}}.
\end{displaymath}
We have assumed that the relative disk half-thickness
$\tilde{H}_{\rm d}$ remains constant with radius for a thin disk,
which is taken as an input parameter in our model. The flux from the
unit surface of a standard thin disk is roughly $f_{\rm
rad}(r)\propto r^{-3}$ \citep*{1973A&A....24..337S}. A pseudo
potential $\Psi_{\rm rad}$ contributed by the radiation force along
the field line is defined as
\begin{equation}
\Psi_{\rm rad}=-{\frac {c}{\kappa_{\rm T}\rho}}\int\mathscr{F}_{{\rm
rad},r}dr-{\frac {c}{\kappa_{\rm T}\rho}}\int\mathscr{F}_{{\rm
rad},z}dz, \label{psi_rad1}
\end{equation}
which are integrated along the field line from the disk surface. The
field line shape is described by Equation (\ref{phi_1}) with a
specified value of $\Phi$. Substitute Equation (\ref{vertical_1})
into Equations (\ref{force_rad_r1}), we can re-write these two
equations in dimensionless form,
\begin{displaymath}
\tilde{\mathscr{F}}_{{\rm rad},r}(r,~z)=\mathscr{F}_{{\rm
rad},r}(r,~z){\frac {cr_{\rm i}^2}{GM\kappa_{\rm T}\rho}}
~~~~~~~~~~~~~~~~~~~~~~~~~~~~~~~~~~~~~~
\end{displaymath}
\begin{equation}
={\frac
{1}{\pi}}\int\limits_{r^\prime}\int\limits_{\phi^\prime}{\frac
{\mu^\prime(1-{\mu^\prime}^2)^{1/2}(r-r^\prime\cos\phi^\prime)\tilde{H}_{\rm
d}}{l(h^2+l^2)(1+\tilde{H}_{\rm d}^2)^{3/2}}}\left({\frac {r_{\rm
i}}{r^\prime}}\right)^3r^\prime dr^\prime d\phi^\prime,
\label{force_rad_r2}
\end{equation}
and
\begin{displaymath}
\tilde{\mathscr{F}}_{{\rm rad},z}(r,~z)=\mathscr{F}_{{\rm
rad},z}(r,~z){\frac {cr_{\rm i}^2}{GM\kappa_{\rm T}\rho}}
~~~~~~~~~~~~~~~~~~~~~~~~~~~~~~~~~~~~~~
\end{displaymath}
\begin{equation}
={\frac
{1}{\pi}}\int\limits_{r^\prime}\int\limits_{\phi^\prime}{\frac
{h^2\tilde{H}_{\rm d}}{(h^2+l^2)^2(1+\tilde{H}_{\rm
d}^2)^{3/2}}}\left({\frac {r_{\rm i}}{r^\prime}}\right)^3r^\prime
dr^\prime d\phi^\prime, \label{force_rad_z2}
\end{equation}
where $r_{\rm i}$ is the footpoint of the field line at the disk
surface. The dimensionless pseudo potential along a magnetic field
line contributed by the radiation force is
\begin{equation}
\tilde{\Psi}_{\rm rad}(\tilde{r},~\tilde{z})=\Psi_{\rm
rad}(r,~z){\frac {r_{\rm i}}{GM}}=-\int\tilde{\mathscr{F}}_{{\rm
rad},r}d\tilde{r}-\int\tilde{\mathscr{F}}_{{\rm rad},z}d\tilde{z},
\label{psi_rad2}
\end{equation}
where $\tilde{r}=r/r_{\rm i}$ and $\tilde{z}=z/r_{\rm i}$.

\subsection{Outflows driven by the magnetic field and radiation
pressure}

An isothermal outflow driven by the magnetic field and radiation
force of the disk along the field line is described by the following
Equations \citep*[cf.][]{1994A&A...287...80C}:
\begin{equation}
p=\rho c_{\rm s,i}^2, \label{c_s_1}
\end{equation}
\begin{equation}
v_{\rm p}={\frac {f}{\rho}}B_{\rm p}, \label{v_p_1}
\end{equation}
\begin{equation}
(v_\phi-\Omega r)B_{\rm p}=v_{\rm p}B_\phi,\label{v_phi_1}
\end{equation}
\begin{equation}
r\left(v_\phi-{\frac {B_{\rm p}B_\phi}{4\pi\rho v_{\rm
p}}}\right)=\Omega r_{\rm A},\label{angular_1}
\end{equation}
and
\begin{equation}
{\frac {v_{\rm p}^2}{2}}+{\frac {1}{2}}(v_\phi-\Omega r)^2+c_{\rm
s,i}^2\ln {\frac {\rho}{\rho_{\rm A}}}+\Psi_{\rm
eff}(r,~z)=E,\label{energy_1}
\end{equation}
where the effective potential along the field line threading the
accretion disk with angular velocity $\Omega$ at the field footpoint
$r_{\rm i}$ is
\begin{equation}
\Psi_{\rm eff}(r,~z)=-{\frac {GM}{(r^2+z^2)^{1/2}}}-{\frac
{1}{2}}\Omega(r_{\rm i})^2r^2
+\Psi_{\rm rad}(r,~z).
 \label{potent_eff_1}
\end{equation}
The last term in Equation (\ref{potent_eff_1}) is contributed by the
radiation force from the disk (see Equation \ref{psi_rad1}).
Equation (\ref{v_p_1}) is the mass conservation of the outflow along
the field line, where $v_{\rm p}$ and $B_{\rm p}$ are the poloidal
velocity and field strength respectively, and $f$ describes the mass
loss rate in the outflow. The conservation of angular momentum is
described by Equation (\ref{angular_1}), in which $r_{\rm A}$ is the
Alfv{\'e}n radius of the outflow. The field line and the motion of
the outflow are parallel in the co-rotating field line frame
(Equation \ref{v_phi_1}). Equation (\ref{energy_1}) is the Bernoulli
Equation of the isothermal outflow.

Substituting Equations (\ref{c_s_1})-(\ref{angular_1}) into Equation
(\ref{energy_1}), we have
\begin{displaymath}
H(r,~\rho,~r_{\rm A},~\rho_{\rm A})={\frac {B_{\rm
p}^2}{8\pi\rho_{\rm A}}}\left({\frac {\rho_{\rm
A}}{\rho}}\right)^2+{\frac {\Omega^2(r_{\rm
A}^2-r^2)^2}{2r^2(1-\rho/\rho_{\rm A})^2}}
\end{displaymath}
\begin{equation}
+c_{\rm s,i}^2\ln{\frac {\rho}{\rho_{\rm A}}}+\Psi_{\rm eff}=E,
\label{energy_2}
\end{equation}
where $c_{\rm s,i}$ is the sound speed of the gas at the disk
surface/the bottom of the outflow, and $\rho_{\rm A}$ is the density
of the outflow at the Alfv{\'e}n radius $r_{\rm A}$. It can be
re-written in dimensionless form as
\begin{displaymath}
\tilde{H}(x,~y,~x_{\rm i},~y_{\rm i})={\frac {r_{\rm
i}}{GM}}H(r,~\rho,~r_{\rm A},~\rho_{\rm A})~~~~~~~~~~~~~~~~~~~~~~~~
\end{displaymath}
\begin{equation}
={\frac {\beta_{\rm i}y_{\rm i}\Theta_{\rm i}}{y^2}}+{\frac
{(1-x^2)^2\tilde{\Omega}^2}{2x^2x_{\rm i}^2(1-y)^2}}+\Theta_{\rm
i}\ln y+\tilde{\Psi}_{\rm eff}=\tilde{E}, \label{energy_3}
\end{equation}
where
\begin{displaymath}
x=r/r_{\rm A};~~~~~y=\rho/\rho_{\rm A};~~~~~\Theta_{\rm i}={\frac
{c_{\rm s,i}^2}{r_{\rm i}^2\Omega_{\rm K}^2}};
\end{displaymath}
\begin{equation}
\beta_{\rm i}={\frac {B_{\rm p,i}^2}{8\pi}}/p_{\rm i}={\frac {B_{\rm
p,i}^2}{8\pi}}/\rho_{\rm i}c_{\rm s,i}^2,\label{beta_i_1}
\end{equation}
and the subscripts ``i" refer to the values of the quantities at the
bottom of the outflow.

Substituting Equation (\ref{vertical_1}) into Equation
(\ref{potent_eff_1}), the effective potential along the field line
threading the accretion disk with angular velocity $\Omega$ at
radius $r_{\rm i}$ can be re-written in dimensionless form,
\begin{equation}
\tilde{\Psi}_{\rm eff}(\tilde{r},~\tilde{z})={\Psi_{\rm
eff}(r,~z)}{\frac {r_{\rm i}}{GM}}=-{\frac
{1}{(\tilde{r}^2+\tilde{z}^2)^{1/2}}}-{\frac
{1}{2}}\tilde{\Omega}^2\tilde{r}^2
+\tilde{\Psi}_{\rm rad}(\tilde{r},~\tilde{z}), \label{potent_eff_2}
\end{equation}
where $\tilde{r}=r/r_{\rm i}$, $\tilde{z}=z/r_{\rm i}$, and the last
term is calculated with Equation (\ref{psi_rad2}).

For a given magnetic field configuration, the outflow along a field
line passes through three critical points, i.e., the slow sonic,
Alfv{\'e}n, and fast sonic points
\citep*[][]{1985A&A...152..121S,1994A&A...287...80C}. At the two
sonic points, the function $\tilde{H}(x,~y,~x_{\rm i},~y_{\rm i})$
satisfies
\begin{equation}
{\frac {\partial \tilde{H}}{\partial x}}={\frac {\partial
\tilde{H}}{\partial y}}=0, \label{phpxy_1}
\end{equation}
and
\begin{equation}
\tilde{H}(x_{\rm s},~y_{\rm s},~x_{\rm i},~y_{\rm
i})=\tilde{H}(x_{\rm f},~y_{\rm f},~x_{\rm i},~y_{\rm i})=\tilde{E}.
\label{hsf_1}
\end{equation}
An additional condition,
\begin{equation}
\tilde{H}(x,~y,~x_{\rm i},~y_{\rm i})=\tilde{E},\label{hi_1}
\end{equation}
is imposed at the disk surface, i.e., $x=x_{\rm i}$, $y=y_{\rm i}$.
A set of seven Equations (\ref{phpxy_1}), (\ref{hsf_1}), and
(\ref{hi_1}), can be solved for seven variables, $x_{\rm s}$,
$y_{\rm s}$, $x_{\rm f}$, $y_{\rm f}$, $x_{\rm i}$, $y_{\rm i}$, and
$\tilde{E}$, for an outflow along a given magnetic field line, when
the temperature and density of the gas at the bottom of the outflow
are specified.

With the derived outflow solution, the mass loss rate in the outflow
from the unit surface area is available,
\begin{equation}
\dot{m}_{\rm w}=\rho_{\rm i}v_{\rm p,i}{\frac {B_z}{B_{\rm
p,i}}}=\rho_{\rm i}v_{\rm p,i}{\frac
{\kappa_0}{(1+\kappa_0^2)^{1/2}}}, \label{mdot_w_1}
\end{equation}
which can be re-written in dimensionless form as
\begin{equation}
{\frac {\dot{m}_{\rm w}}{\Sigma_{\rm d}\Omega_{\rm K}}}={\frac
{\beta\tilde{H}_{\rm d}(1+\kappa_0^2)^{1/2}}{\kappa_0[2\beta_{\rm
i}\Theta_{\rm i}y_{\rm i}(1+\tilde{H}_{\rm d}^2)^3]^{1/2}}}.
\label{mdot_w_2}
\end{equation}
Here, Equations (\ref{vertical_1}), (\ref{p_d_1}), (\ref{f_rad_1})
and (\ref{energy_3}) are used. The reciprocal of $\dot{m}_{\rm
w}/\Sigma_{\rm d}\Omega_{\rm K}$ represents the number of orbits in
which all the gas in the disk is channeled into the outflow.

\subsection{Boundary conditions}

The solution of an outflow along a given field line is available by
solving a set of seven non-linear algebraic equations with suitable
boundary conditions at the disk surface. The dimensionless
temperature $\Theta_{\rm i}$ of the gas is an input parameter, and
the density of the gas is described by $\beta_{\rm i}$ if the value
of $\beta$ is specified (see Equation \ref{beta_i_1}). For the
outflow can be efficiently driven by the magnetic field, the gas
pressure should be lower than the magnetic pressure, i.e.,
$\beta_{\rm i}\ga 1$ is required. For most calculations in this
work, $\beta_{\rm i}=1$ is used.

The temperature of the photosphere roughly equals to the surface
temperature of the disk, $T_{\rm i}$, which is given by
\begin{equation}
T_{\rm i}={\frac {f_{\rm rad}^{1/4}}{\sigma^{1/4}}}. \label{t_s}
\end{equation}
The sound speed of the gas in the photosphere of the disk is
\begin{equation}
c_{\rm s,i}=\left({\frac {kT_{\rm i}}{\mu m_{\rm p}}}\right)^{1/2}.
\label{c_s1}
\end{equation}
Substituting Equations (\ref{vertical_1}) and (\ref{t_s}) into
Equation (\ref{c_s1}), we have
\begin{displaymath}
{\frac {c_{\rm s,i}^2}{r_{\rm i}^2\Omega_{\rm K}^2}}
=\left[(1.895M_\odot)^2{\frac {60G^3}{\pi^2\kappa_{\rm T}}}{\frac
{r_{\rm i}\Omega_{\rm K}^2\tilde{H}_{\rm d}}{(1+\tilde{H}_{\rm
d}^2)^{3/2}}} \right]^{1/4}{\frac {r_{\rm i}}{GM\mu}}
\end{displaymath}
\begin{equation}
=4.946\times10^{-6}\mu^{-1}(r_{\rm i}/r_{\rm S})^{1/2}\tilde{H}_{\rm
d}^{1/4}m^{-1/4}(1+\tilde{H}_{\rm d}^2)^{-3/8},
 \label{c_s2}
 \end{equation}
where
\begin{equation}
r_{\rm S}={\frac {2GM}{c^2}},~~~~~~~~m={\frac
{M}{M_\odot}},~~~~~~~~~~~~~~~~~~~~~~~~~
\end{equation}
and $(\hbar{c}/G)^{3/2}m_{\rm p}^{-2}=1.895 M_\odot$ is used. The
molecular weight $\mu=0.5$ for pure hydrogen plasma.

The surface temperature of a gas pressure dominated accretion disk
is
\begin{equation}
{\frac {c_{\rm s,i}^2}{r_{\rm i}^2\Omega_{\rm K}^2}}=\left({\frac
{4}{3\tau}}\right)^{1/4}{\frac {c_{\rm s,c}^2}{r_{\rm
i}^2\Omega_{\rm K}^2}}=\left({\frac {4}{3\tau}}
\right)^{1/4}\tilde{H}_{\rm d}^2, \label{c_s_gpd1}
\end{equation}
where $c_{\rm s,c}$ is the sound speed of the gas at the mid-plane
of the disk, and $\tau$ is the optical depth of the disk in the
vertical direction. The optical depth $\tau$ is in the range of
$\sim 10^2-10^5$ for a thin accretion disk accreting at different
rates, which means the surface temperature of the disk, ${c_{\rm
s,i}^2}/{r_{\rm i}^2\Omega_{\rm K}^2}\sim 0.06-0.34\tilde{H}_{\rm
d}^2$ \citep*[][]{2013ApJ...765..149C}. Compared with a gas pressure
dominated accretion disk, the surface temperature of a radiation
pressure dominated accretion disk is much lower, ${c_{\rm
s,i}^2}/{r_{\rm i}^2\Omega_{\rm K}^2}\ll \tilde{H}_{\rm d}^2$ (see
Equation \ref{c_s2}). The accretion disk is gas pressure dominated
in the outer region or/and its mass accretion rate is low
\citep*[][]{1989MNRAS.238..897L,1973A&A....24..337S}. This implies
gas pressure gradient in the outflow from the inner region of a
luminous (radiation pressure dominated) disk is almost negligible,
and therefore the cold gas approximation is a good approximation
\citep*[][]{2012MNRAS.426.2813C}. In this case, the outflow will be
suppressed if the field line is inclined at an angle slightly larger
than the critical one, because the internal energy of such cold gas
at the disk surface is too low to overcome the effective potential
barrier along the field line. If the angle is lower than the
critical one, the slow sonic point may probably go into the disk,
and the density of the gas in the disk is usually high, which
usually leads to a slow dense circular flow above the disk
\citep*[see][for the detailed
discussion]{1994A&A...287...80C,1996epbs.conf..249S}. It implies
that hot gas is needed for feeding an outflow from a radiation
pressure dominated accretion disk.

The observed ultraviolet (UV)/optical emission of AGN is thought to
be the thermal emission from the standard geometrically thin,
optically thick accretion disks
\citep*[e.g.,][]{1978Natur.272..706S,1982ApJ...254...22M,1989ApJ...346...68S}.
The observed power-law hard X-ray spectra of AGN are most likely due
to the inverse Compton scattering of soft photons on a population of
hot electrons in the coronas above the disk
\citep*[][]{1979ApJ...229..318G,1993ApJ...413..507H,1994ApJ...432L..95H}.
It has been found that the temperature of the hot electrons in the
corona is roughly around 10$^9$~K, which can successfully reproduce
a power-law hard X-ray spectrum as observed
\citep*[e.g.,][]{2003ApJ...587..571L,2009MNRAS.394..207C}. In this
work, we consider a hot corona above a thin accretion disk, which is
a reservoir to supply hot gas to feed the outflow driven by the
magnetic field and radiation force of the disk. The detailed
properties of the corona are still quite unclear. For simplicity, we
use two parameters, the temperature $\Theta_{\rm i}$ and the density
of the corona, in our model calculations. As the magnetic field
strength at the disk surface is described by the value of $\beta$
specified, the density of the corona is given by the ratio of
magnetic to gas pressure in the corona ($\beta_{\rm i}$) for given
temperature $\Theta_{\rm i}$ and disk field strength $\beta$ (see
Equation \ref{beta_i_1}).

\section{Results}

We compare the temperatures of the photospheres between the
radiation pressure dominated disks and gas pressure dominated
accretion disks in Figure \ref{cs_i}. As discussed in Section 2.5,
the temperature of the photosphere of a radiation pressure dominated
accretion disk is indeed much lower than that of a gas pressure
dominated accretion disk. We note that the surface temperature of
the radiation pressure dominated disk decreases with increasing
black hole mass.

We use the magnetic field configuration above the disk given in
Section 2.2, of which the field line is given by setting
$\Phi=const$. In our calculations, the relative disk thickness
$\tilde{H}_{\rm d}$ is assumed to be independent of radius, which is
an input parameter. The angular velocity of the disk can be
calculated with Equation (\ref{Omega_2}), if the strength of the
magnetic field $\beta$ at the disk surface, the disk thickness
$\tilde{H}_{\rm d}$, and the field line inclination $\kappa_0$ are
specified. The field line inclination $\kappa_0$ at the disk surface
is a function of $r/r_{\rm 0}$.

With the specified values of $\Theta_{\rm i}$ and $\beta_{\rm i}$ of
the corona, we can calculate the dynamics of the outflow along a
given magnetic field line as described in Section 2. In Figures
\ref{b_conf_h0p1th0p01}-\ref{b_conf_h0p1bt0p75mr}, we plot the
location of the critical points in the outflow solutions derived
with different values of the model parameters. We show how the
results vary with the magnetic field strength $\beta$ in Figure
\ref{b_conf_h0p1th0p01}. The rotational velocity $\tilde{\Omega}$ of
the disk decreases with increasing field strength $\beta$ (see
Equation \ref{Omega_2}), and the effective potential barrier
increases with decreasing $\tilde{\Omega}$. Thus, the critical
points go far away from the disk. In Figure \ref{b_conf_h0p1bt0p75},
we plot the results calculated with different values of corona
temperature $\Theta_{\rm i}$.
In order to show the role of the radiation
force on the acceleration of the outflows, we compare the results
with those of the purely magnetically driven outflows in Figure
\ref{b_conf_h0p1bt0p75mr}. The location of the critical points in
the pure magnetically driven outflows is farther away from the disk
compared with those of the outflows driven by the magnetic field
together with the radiation force. This is because the effective
potential barrier becomes deep for a pure magnetically driven
outflow (see Figure \ref{psi_eff_h0p1}).

We can calculate the mass loss rate in the outflow with Equation
(\ref{mdot_w_2}) when the outflow solution is derived. In Figures
\ref{wind_mdot}, we plot the mass loss rates as functions of the
field line inclination $\kappa_0$ at the disk surface. The
velocities of the outflows along the field lines are given in
Figures \ref{vp_vphi} and \ref{vp_vphib}. The mass loss rates
decrease with $\kappa_0$, while the velocities of the outflows
increase with the field line inclination $\kappa_0$ (see Figures
\ref{wind_mdot}-\ref{vp_vphib}). We find that the dynamical
properties of the outflows are different near the disk surface,
while the poloidal velocities converge at large distances along the
field lines with the same $\kappa_0$ for low-$\kappa_0$ case. It is
found that the velocities of the flows increase with decreasing
density of the corona (large-$\beta_{\rm i}$), if the values of all
the other parameters are fixed (see Figures \ref{vp_vphi} and
\ref{vp_vphib}).


 \figurenum{1}
\centerline{\includegraphics[angle=0,width=7.5cm]{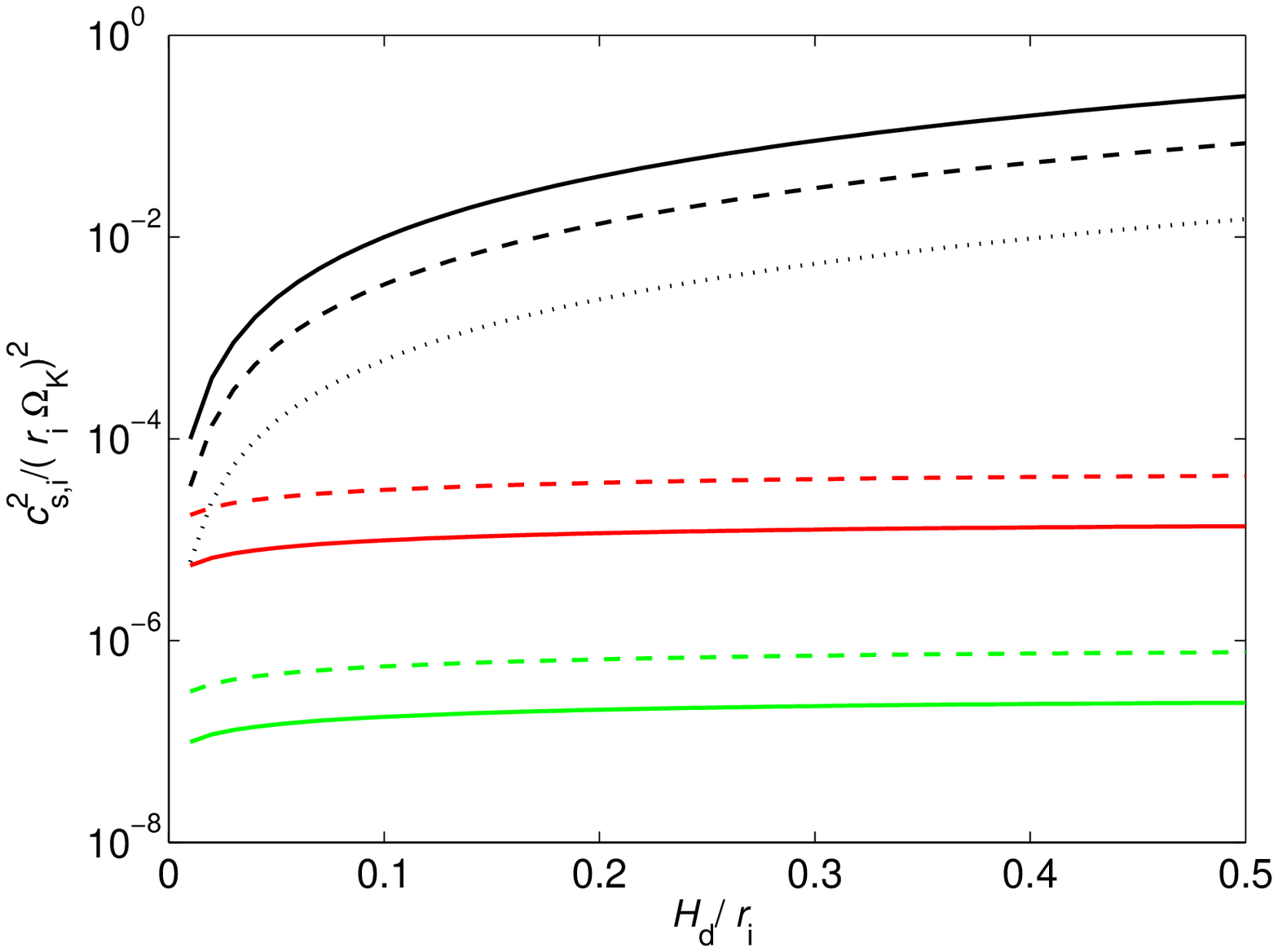}}
\figcaption{The temperature of the photosphere of the disk. The
coloured lines represent the results for radiation pressure
dominated accretion disks (see Section 2.5 for the detailed
discussion). The red lines are the results calculated with a black
hole mass $M=10M_\odot$, while the green lines are for a massive
black hole with $M=10^8M_\odot$. The solid lines indicate the
temperature at radius $r=10r_{\rm S}$, and the dashed lines are for
$r=100r_{\rm S}$. The black solid line represent the temperature of
the photosphere of an isothermal gas pressure dominated accretion
disk, while the dashed and dotted lines are the results calculated
with $\tau=100$ and $10^5$, respectively. \label{cs_i}
}\centerline{}


 \figurenum{2}
\centerline{\includegraphics[angle=0,width=7.5cm]{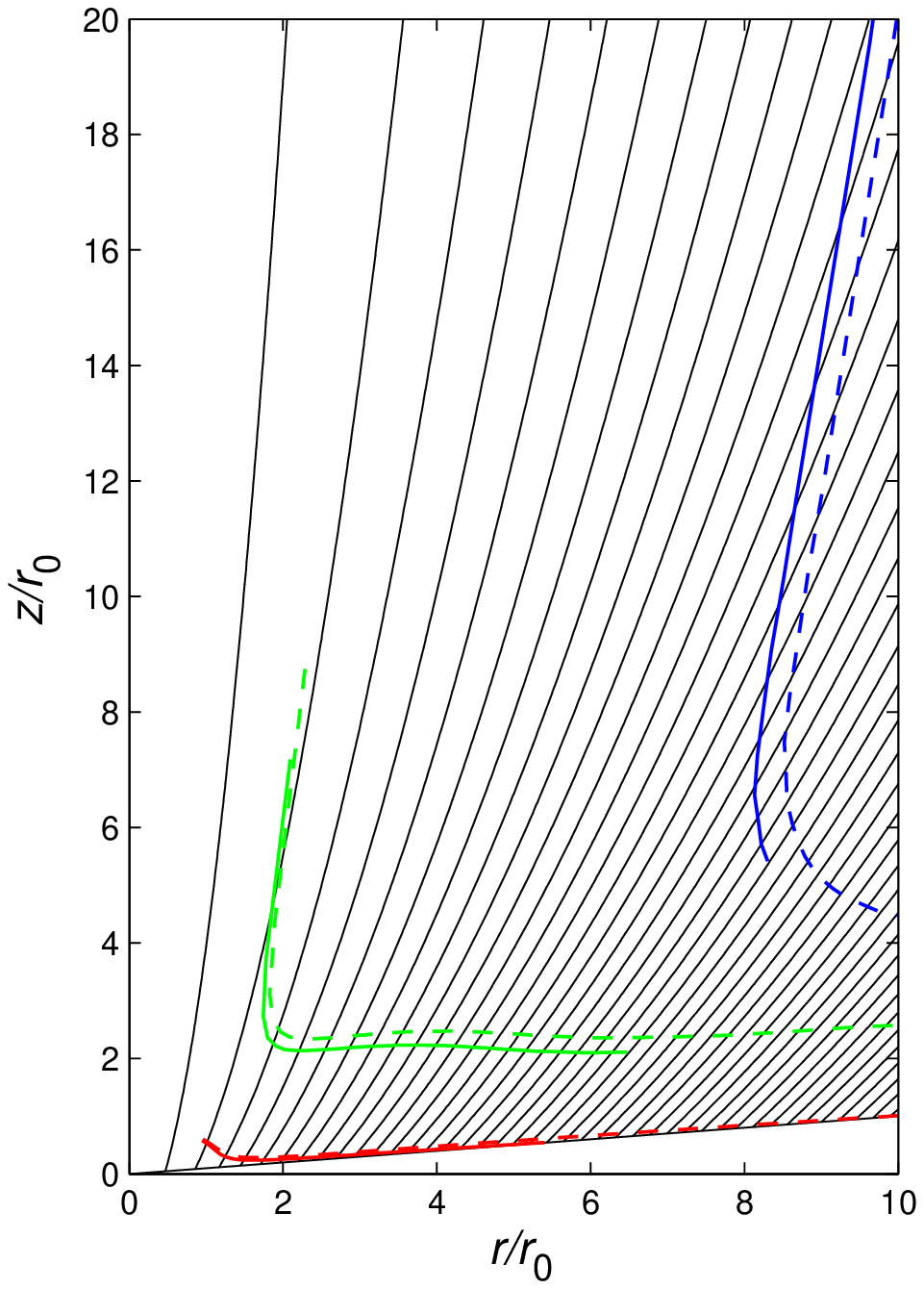}}
\figcaption{The locations of the critical points of the outflow
solutions (red: slow sonic points; green: Alfv{\'e}n points; blue:
fast sonic points). The disk thickness $\tilde{H}_{\rm d}=0.1$, the
dimensionless temperature of the hot corona $\Theta_{\rm i}=0.01$,
and the ratio of magnetic to gas pressure in the corona $\beta_{\rm
i}=1$, are adopted in the calculations. Different types of lines
represent the results for different values of $\beta=0.75$ (solid)
and $1$ (dashed), respectively. \label{b_conf_h0p1th0p01}
}\centerline{}


 \figurenum{3}
\centerline{\includegraphics[angle=0,width=7.5cm]{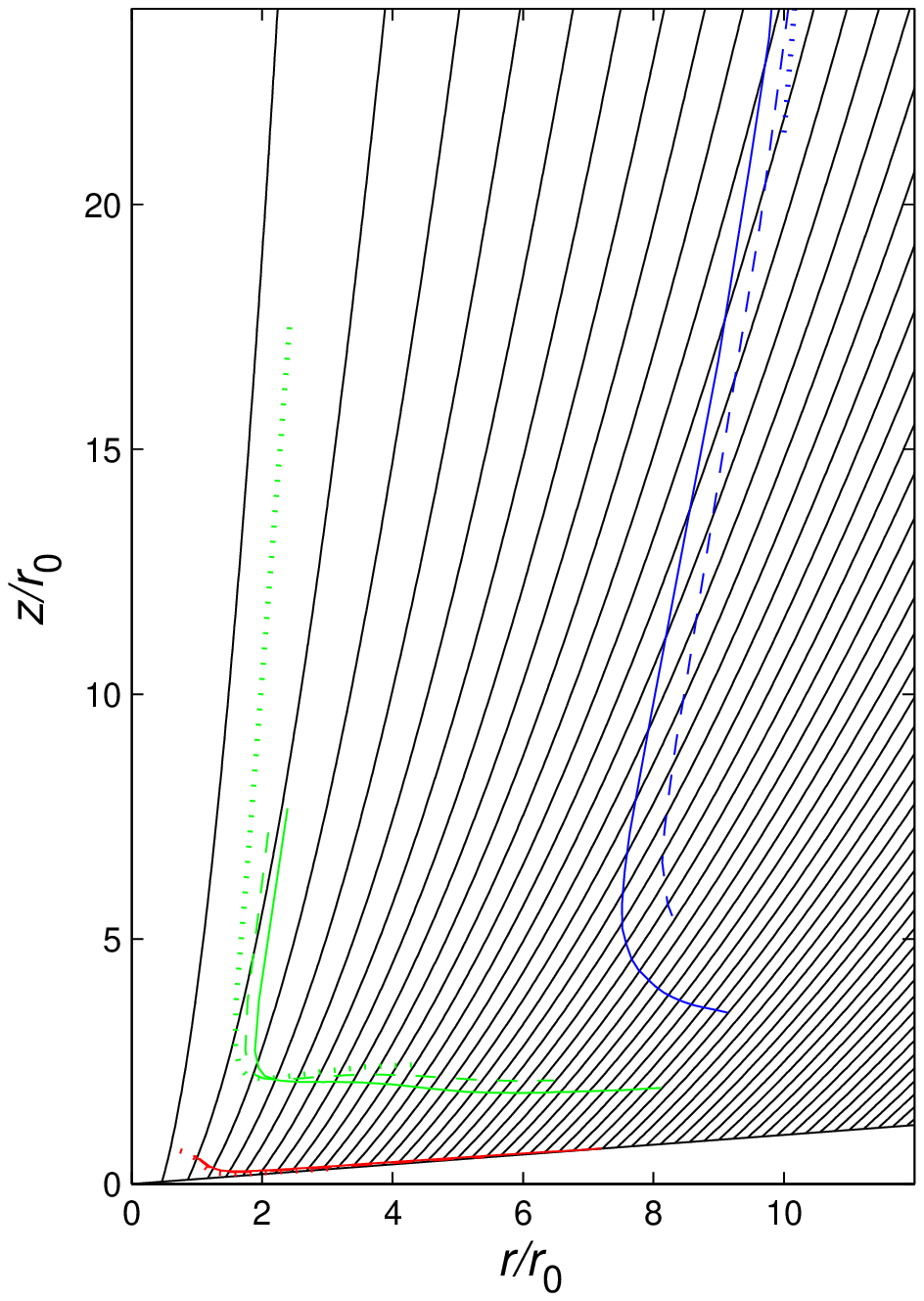}}
\figcaption{The same as Figure \ref{b_conf_h0p1th0p01}. In all
calculations, $\beta=0.75$ is adopted. Different types of lines
represent the results for different values of $\Theta_{\rm i}=0.005$
(solid), $0.01$ (dashed), and $0.02$ (dotted), respectively.
\label{b_conf_h0p1bt0p75} }\centerline{}


 \figurenum{4}
\centerline{\includegraphics[angle=0,width=7.5cm]{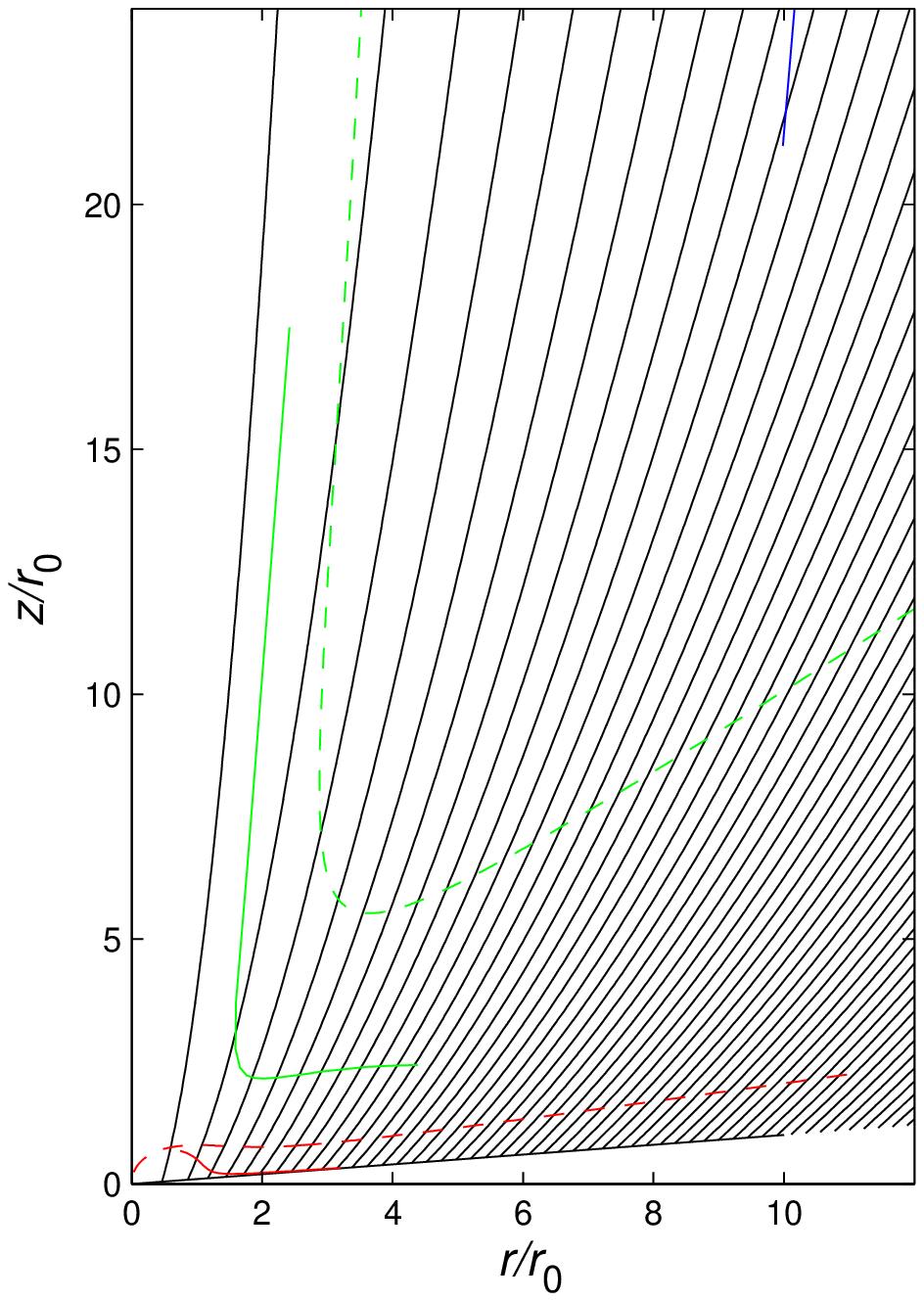}}
\figcaption{The locations of the critical points of the outflow
solutions (red: slow sonic points; green: Alfv{\'e}n points; blue:
fast sonic points). The disk thickness $\tilde{H}_{\rm d}=0.1$, the
dimensionless temperature of the hot corona $\Theta_{\rm i}=0.02$,
and the ratio of magnetic to gas pressure in the corona $\beta_{\rm
i}=1$, and $\beta=0.75$ are adopted in the calculations. For
comparison, the results without considering the effects of radiation
pressure are also plotted (dashed lines).
\label{b_conf_h0p1bt0p75mr} }\centerline{}


 \figurenum{5}
\centerline{\includegraphics[angle=0,width=7.5cm]{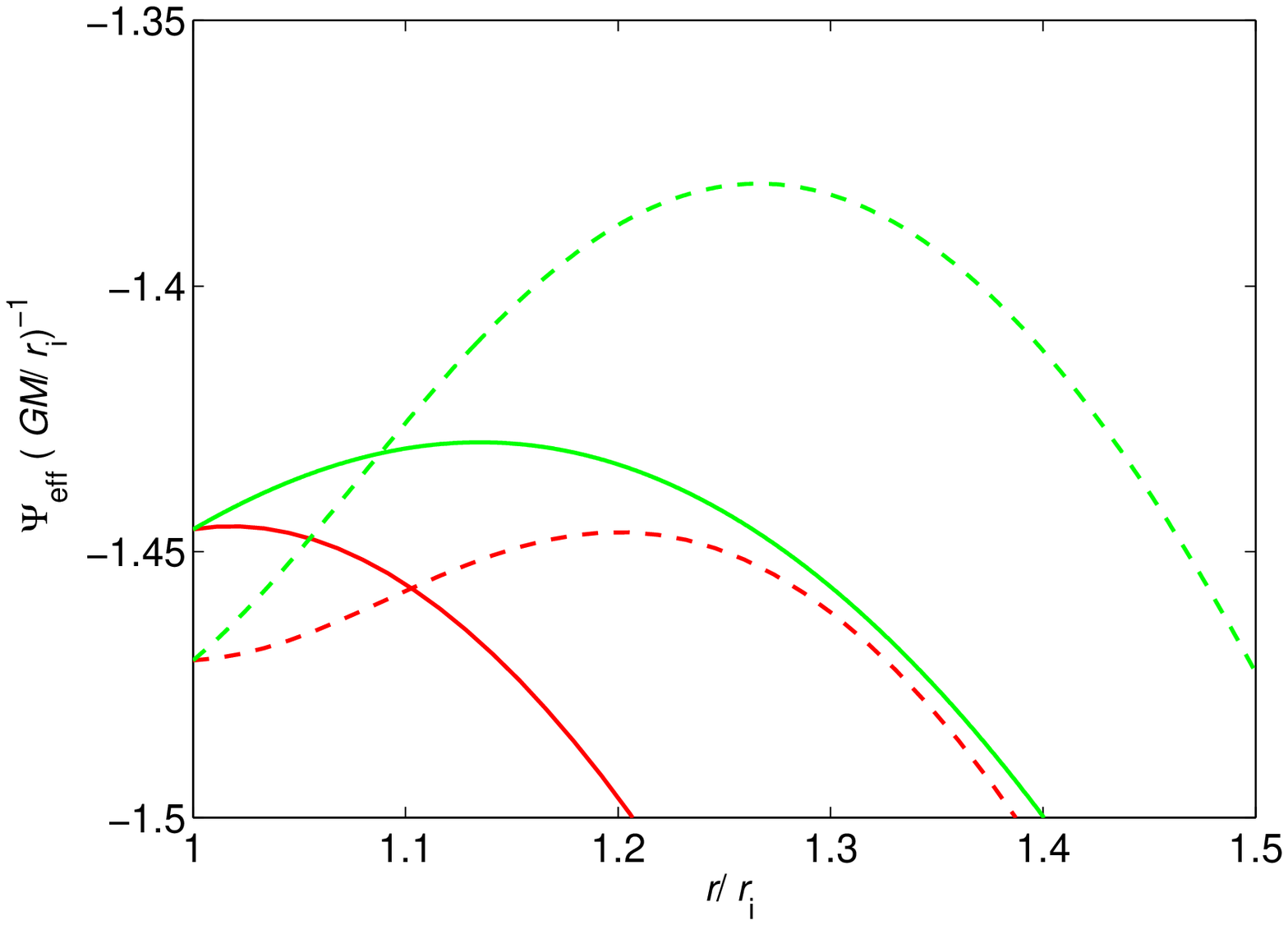}}
\figcaption{The effective potential along the field line threading a
disk with $\tilde{H}_{\rm d}=0.1$ and $\beta=0.75$. The solid lines
represent the results for the field line inclination $\kappa_0=1.5$
at the disk surface, while the dashed lines are for $\kappa_0=3$.
For comparison, we plot the results calculated without considering
radiation pressure effects as green lines. \label{psi_eff_h0p1}
}\centerline{}


 \figurenum{6}
\centerline{\includegraphics[angle=0,width=7.5cm]{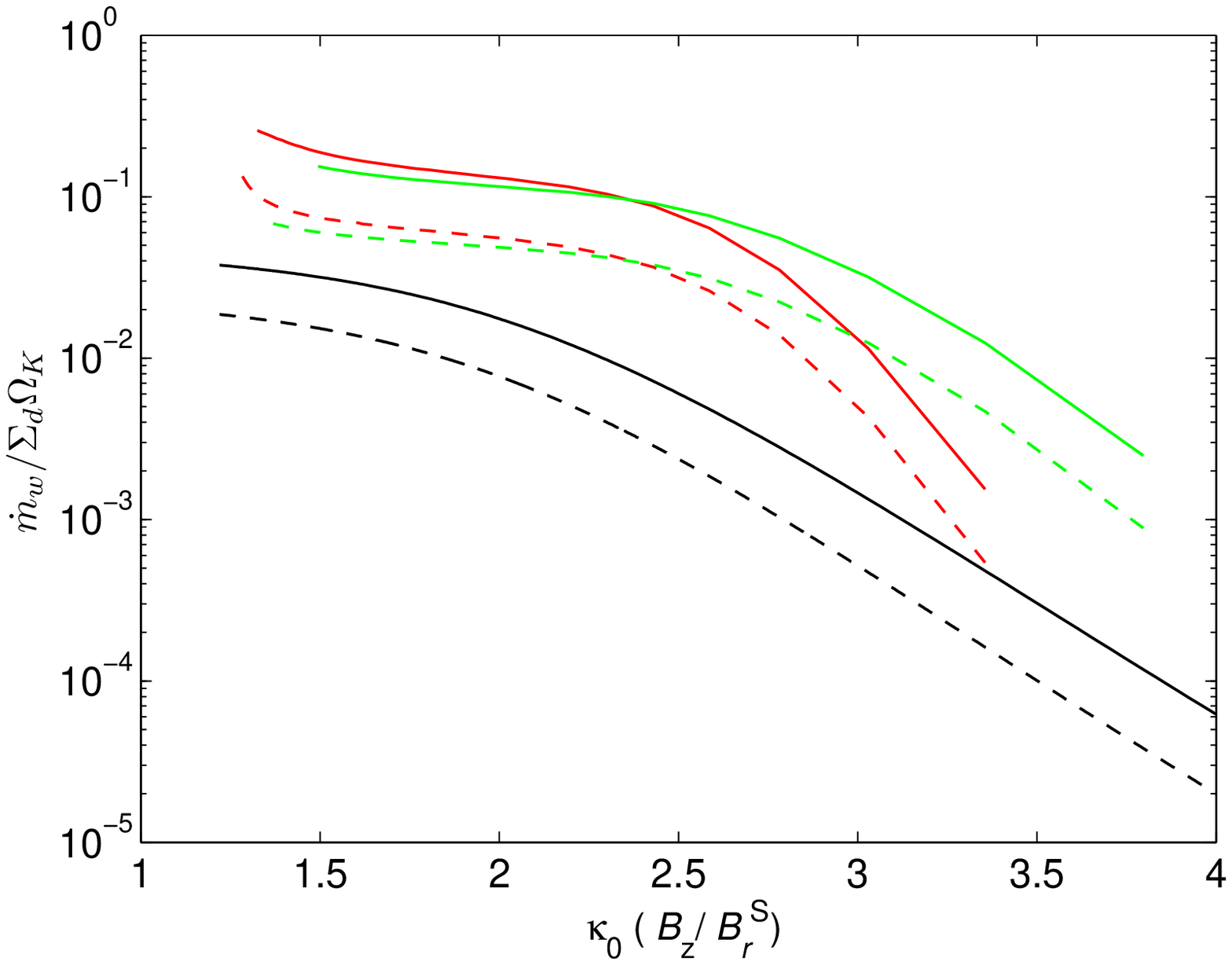}}
\figcaption{The dimensionless mass loss rates in the outflows as
functions of field line inclination $\kappa_0=B_z/B_r^{\rm S}$ at
the disk surface. In all the calculations, $\beta=0.75$ and
$\tilde{H}_{\rm d}=0.1$ are adopted. The solid lines represent the
results calculated with $\beta_{\rm i}=1.0$, while the dashed lines
are for $\beta_{\rm i}=3$. The different colors are for different
corona temperature, $\Theta_{\rm i}=0.01$ (red) and $0.02$ (green).
For comparison, we plot the result for purely magnetically driven
outflows with $\Theta_{\rm i}=0.02$ as black lines (solid line:
$\beta_{\rm i}=1$, and dashed lines: $\beta_{\rm i}=3$).
\label{wind_mdot} }\centerline{}


 \figurenum{7}
\centerline{\includegraphics[angle=0,width=7.5cm]{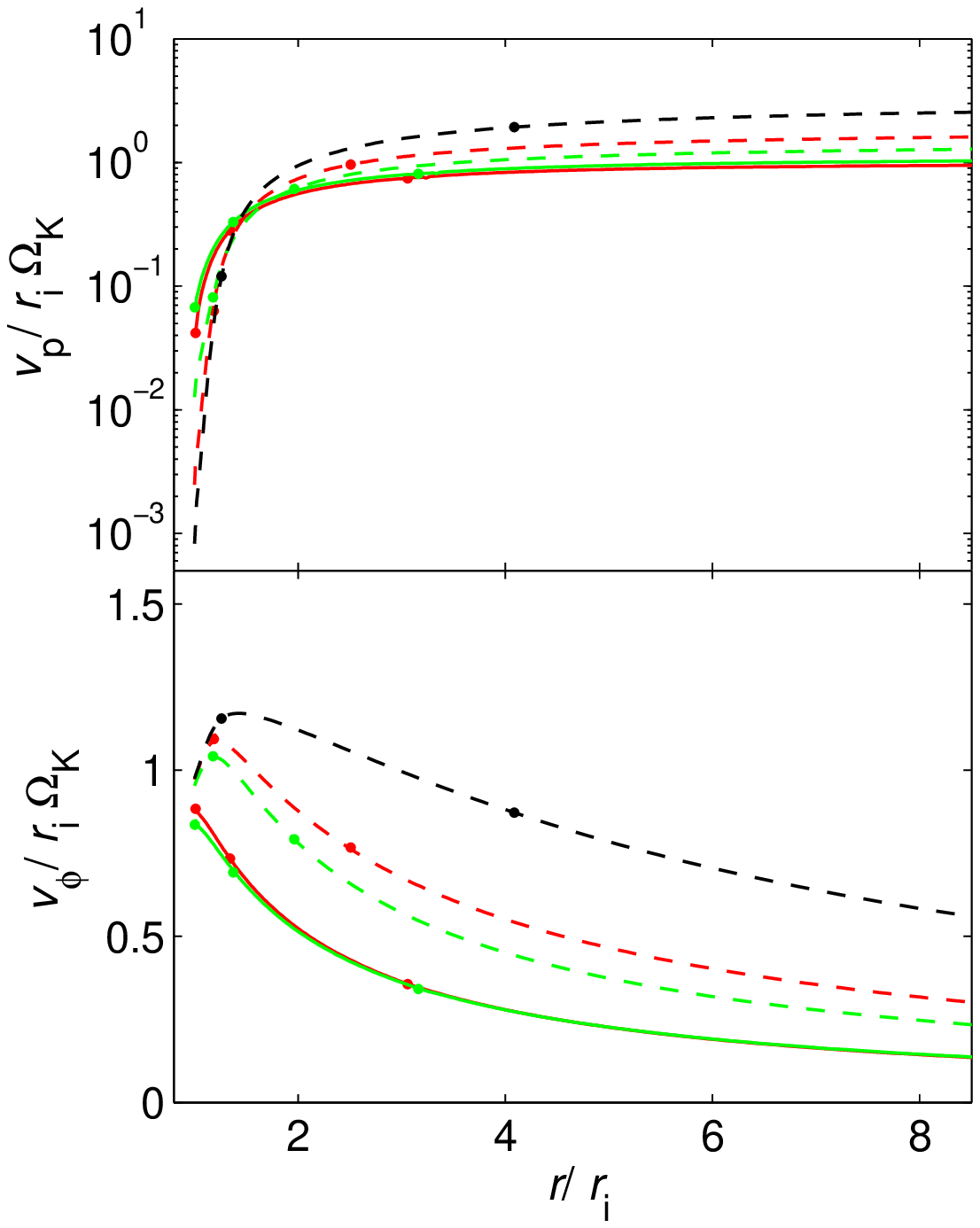}}
\figcaption{The poloidal and azimuthal velocities of the outflow
along the field line. The ratio of magnetic to gas pressure in the
corona $\beta_{\rm i}=1$, and $\beta=0.75$, are adopted in all the
calculations. The solid lines represent the outflow along the field
line with $\kappa_0=1.5$ at the disk surface, while the dashed lines
are for $\kappa_0=3$. The colored lines represent for the results
with different values of model parameters (red lines: $\Theta_{\rm
i}=0.01$, and green lines: $\Theta_{\rm i}=0.02$). The black dashed
lines are the results calculated with pure magnetically driven
outflow model ($\kappa_0=3$ and $\Theta_{\rm i}=0.02$). The dots
indicate the critical points in the outflow, i.e., slow sonic,
Alfv{\'e}n, and fast sonic points (from left to right).
\label{vp_vphi} }\centerline{}


 \figurenum{8}
\centerline{\includegraphics[angle=0,width=7.5cm]{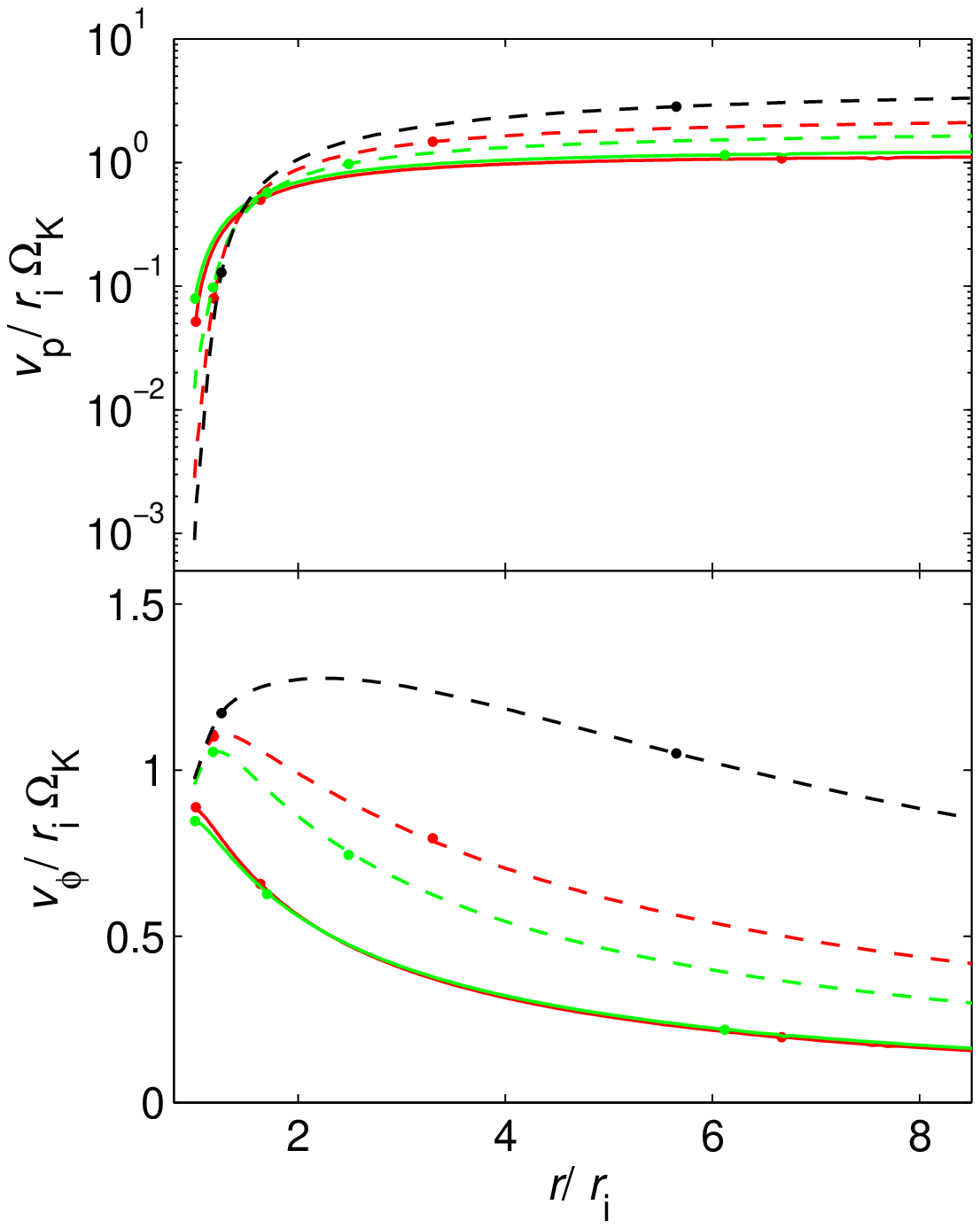}}
\figcaption{The same as Figure \ref{vp_vphi}, but $\beta_{\rm i}=3$
is adopted. \label{vp_vphib} }\centerline{}

\section{Discussion}

We find that the temperature of the photosphere at the surface of a
radiation pressure disk is significantly lower than that for a gas
pressure dominated accretion disk with the same relative disk
thickness $\tilde{H}_{\rm d}$ (see Figure \ref{cs_i}). The surface
temperature of the radiation pressure disk is insensitive to the
disk thickness $H_{\rm d}/r$, while it decreases with increasing
black hole mass. The gas pressure gradient may help accelerating the
outflow. The importance of the gas pressure on driving the outflow
can be estimated by comparing $c_{\rm s}^2$ with the effective
potential barrier $\Delta \Psi_{\rm eff}=\Psi_{\rm eff}^{\rm
max}-\Psi_{\rm eff,i}$ along the field line. In the magnetically
driven outflow model, the mass loss rate in the outflow
$\dot{m}_{\rm w}\propto\exp(-\Delta \Psi_{\rm eff}/c_{\rm s}^2)$
\citep*[see][for the
details]{1996epbs.conf..249S,1998ApJ...499..329O}. For such a
low-temperature photosphere at the surface of the radiation pressure
dominated disk, an extremely shallow potential barrier
(small-$\Delta \Psi_{\rm eff}$) is required for launching an
outflow, which means the values of $\beta$ and $\kappa_0$ should be
in the narrow ranges close to those of the cold gas launching
condition given in \citet{2012MNRAS.426.2813C}. It implies strict
constraints on the outflows that can be driven from the disk
surface, especially for massive black hole cases, as the
temperatures of the spheres are above two orders of magnitude lower
than those of stellar black hole accretion disks (see Figure
\ref{cs_i}). The outflow can be launched along the field line if the
effective potential $\Psi_{\rm eff}$ monotonically decreases along
the field line. In this case, the outflow is alternatively
accelerated within the disk, which is, for example, similar to the
solutions of the outflows along the field lines inclined at small
angles with respect to the disk surface given in
\citet{1994A&A...287...80C}. The gas overcomes a potential barrier
along the field line in the disk, and the slow sonic point is
located within the disk. Thus, the slow dense circular flows are
usually present above the disks in this case
\citep*[][]{1994A&A...287...80C}. Such calculations can be done with
a known field configuration within the disk, which is beyond the
scope of this work, and we will not investigate this kind of
solutions in this paper.

As discussed in Section 2.5, the outflow can be driven from the hot
gas in the corona above the disk. The field configuration can be
derived with the detailed physics of advection and diffusion of
magnetic fields considered
\citep*[][]{1989ASSL..156...99V,1994MNRAS.267..235L,2009ApJ...701..885L,2012MNRAS.424.2097G,2013MNRAS.430..822G,2013ApJ...765..149C},
which is beyond the scope of this work. We use the configuration of
the field given in Section 2.2, of which the strength at the disk
surface decreases with increasing radius. As we focus on how the hot
gas at the disk surface is accelerated along the field line with the
help with the radiation force of the disk, the above mentioned field
configuration is sufficient for our present investigation.

The location of the critical points of the outflow solutions with
different values of the accretion disk/corona parameters is plotted
in Figures \ref{b_conf_h0p1th0p01}-\ref{b_conf_h0p1bt0p75mr}. It is
found that the slow sonic points are always close to the disk
surface. In the central region of the disk, the field lines are
inclined at large angles with respect to the disk surface. The mass
loss rate in the outflow decreases with increasing inclination
$\kappa_0$ (see Figure \ref{wind_mdot}), because of the effective
potential barrier increases with $\kappa_0$ (see Figure
\ref{psi_eff_h0p1}) and $\dot{m}_{\rm w}\propto\exp(-\Delta
\Psi_{\rm eff}/c_{\rm s}^2)$. The Alfv{\'e}n and fast sonic points
go farther away from the disk when $\kappa_0$ is large (see Figures
\ref{b_conf_h0p1th0p01}-\ref{b_conf_h0p1bt0p75mr}), which is caused
by a tenuous low-$\dot{m}_{\rm w}$ outflow. The mass loss rate in
the outflow increases with the temperature $\Theta_{\rm i}$ of the
gas in the corona, if the values of all other parameters are fixed.
In order to illustrate the role of the
radiation force on the acceleration of the outflow, we compare the
results of pure magnetically driven outflows in Figure
\ref{wind_mdot}. The mass loss rates in the outflows driven by the
magnetic field and radiation pressure are significantly higher than
those for pure magnetically driven outflows, because the radiation
pressure makes the effective potential barrier shallow.

Less mass is loaded in the outflow in high-$\kappa_0$ cases, and the
outflow can be driven to a relative high velocity, while a slow
dense outflow is present when $\kappa_0$ is small. It is found that
the poloidal velocities of the outflows almost converge at large
distances along the field line with a given $\kappa_0$ for
low-$\kappa_0$ cases. This implies that the gas in the dense
outflows is dominantly accelerated by the radiation pressure of the
disk, which is almost independent of the values of the disk
parameters. However, the situation is different for the azimuthal
velocity of the outflows, which is independent of the radiation
pressure. We find that the outflows with low mass loss rates, i.e.,
low-$\Theta_{\rm i}$, or/and high-$\kappa_0$ as discussed above, can
be accelerated to a large velocity in the azimuthal direction.

\section{Summary}

We estimate the temperature of the photosphere above the radiation
dominated accretion disk, and find that $\Theta\ll \tilde{H}_{\rm
d}^2$. Therefore, it is difficult for the outflows driven from the
photospheres of the radiation dominated disks. This implies that hot
gas (probably in the corona) is necessary for launching an outflow
from the radiation pressure dominated disk, which provides a natural
explanation on the observational evidence that the relativistic jets
are related to hot plasma in some X-ray binaries and active galactic
nuclei \citep*[][]{2011MNRAS.416.1324Z,2013ApJ...770...31W}.

We investigate the outflows accelerated from the hot corona above
the disk by the magnetic field and radiation force of the accretion
disk, and find that the outflow can be driven from the corona with
the help of radiation force even if the field line is inclined at a
large angle ($>60^\circ$) with respect to the disk surface. The
potential barrier decreases due to the radiation force of the disk,
and therefore the mass loss rate in the outflow increases. We find
that slow outflows with high mass loss rates are present if the
field line inclination $\kappa_0\la 2$. This may be the reason why
the jets in radio-loud narrow-line Seyfert galaxies are in general
mild relativistic compared with those in blazars
\citep*[][]{2006AJ....132..531K,2006PASJ...58..777D,2010AJ....139.2612G,2011ApJ...738..126D}.

\acknowledgments This work is supported by the NSFC (grants
11173043, 11121062, and 11233006), the CAS/SAFEA International
Partnership Program for Creative Research Teams (KJCX2-YW-T23), and
Shanghai Municipality.

\end{document}